\documentclass[aps,12pt,a4paper]{revtex4-2}
\usepackage{epsfig}
\usepackage{graphicx}
\usepackage{appendix}
\usepackage{amsmath,amssymb,color}
\usepackage[english]{babel}
\usepackage{hyperref}
\usepackage{natbib}
\usepackage{appendix}
\hypersetup{
    colorlinks=true,
    linkcolor=blue,
    filecolor=magenta,      
    urlcolor=cyan,
    pdftitle={Overleaf Example},
    pdfpagemode=FullScreen,
    }

\parskip=\medskipamount

\newcommand{\eq}[1]{(\ref{#1})}
\newcommand{\fig}[1]{Fig.~\ref{#1}}

\newcommand{\be}{\begin{equation}}
\newcommand{\ee}{\end{equation}}

\newcommand{\la}{\left<}
\newcommand{\ra}{\right>}

\begin{document}
\title{Metric structural human connectomes: localization and multifractality of eigenmodes}

\author{Anna Bobyleva$^1$, Alexander Gorsky$^{2,3}$, Sergei Nechaev$^4$, Olga Valba$^{6,3}$ and Nikita Pospelov$^5$}

\affiliation{$^1$ Department of Biophysics, Faculty of Biology of the Moscow State University, Moscow, Russia \\
$^2$Institute for Information Transmission Problems RAS, 127051 Moscow, Russia \\ 
$^3$Laboratory of Complex Networks, Center for Neurophysics and Neuromorphic Technologies, Moscow, Russia \\
$^4$LPTMS, CNRS -- Universit\'e Paris Saclay, 91405 Orsay Cedex, France \\
$^5$Institute for Advanced Brain Studies of the Moscow State University, Moscow, Russia \\
$^6$Higher School of Economics, Moscow, Russia}


\begin{abstract}

In this study, we explore the fundamental principles behind the architecture of the human brain's structural connectome, from the perspective of spectral analysis of Laplacian and adjacency matrices. Building on the idea that the brain strikes a balance between efficient information processing and minimizing wiring costs, we aim to understand the impact of the metric properties of the connectome and how they relate to the existence of an inherent scale. We demonstrate that a simple generative model, combining nonlinear preferential attachment with an exponential penalty for spatial distance between nodes, can effectively reproduce several key characteristics of the human connectome, including spectral density, edge length distribution, eigenmode localization and local clustering properties. We also delve into the finer spectral properties of the human structural connectomes by evaluating the inverse participation ratios ($\text{IPR}_q$) across various parts of the spectrum. Our analysis reveals that the level statistics in the soft cluster region of the Laplacian spectrum deviate from a purely Poisson distribution due to interactions between clusters. Additionally, we identified scar-like localized modes with large IPR values in the continuum spectrum. We identify multiple fractal eigenmodes distributed across different parts of the spectrum, evaluate their fractal dimensions and find a power-law relationship in the return probability, which is a hallmark of critical behavior. We discuss the conjectures that a brain operates in the Griffiths or multifractal phases.

\end{abstract}

\maketitle

\section{Introduction}

Understanding the underlying principles governing brain structural organization, which provide effective signal propagation, is undoubtedly a challenging and highly demanded task. Presently, the prevailing paradigm suggests that relatively short correlations are dominant, and spatiotemporal processes in the brain can be described in terms of wave dynamics within an underlying graph structure. In the frameworks of this paradigm the key object is the graph Laplacian of the structural connectome, which describes the diffusion-like flow of the information in the brain. The information flow along the structural backbone of the connectome can be connected with the functional brain dynamics via the so-called ''mass model'' with activation-inhibition balance -- see \cite{deco2008dynamic, sporns2013human} for the review.

It is widely accepted that the spectral statistics of the Laplacian provides crucial insight into the structure of the brain \cite{bullmore2009complex, fornito2016fundamentals}, establishing a connection between the spectrum of the structural connectome and cognitive functions. Recent developments in this area further underline the significance of this relationship \cite{luppi2023distributed, aqil2021graph, tewarie2020mapping, abdelnour2018functional, PhysRevE.104.034411}.

Besides, there is a growing body of literature emphasizing the importance of metric geometry of the brain cortex. For instance, in \cite{robinson2016eigenmodes, gabay2017cortical}, it is suggested that brain dynamics are governed by a metric Laplacian, which captures the underlying geometry of the 2D brain surface. These geometric models conceptualize the brain surface as a smooth object that can be effectively described using the conventional continuum metric-dependent Laplacian. Moreover, in models of exponential networks such as those discussed in \cite{robinson2019physical, wang2016brain, roberts2017consistency}, the typical spatial length scale is introduced through the postulation of a cut-off, indicating the presence of a finite correlation length. This implies that understanding the geometry of the brain cortex, as represented by its metric properties, is essential for comprehending the dynamics of neural activity within the brain.

Recently, it has become apparent that geometric models outperform other methods in analyzing functional magnetic resonance imaging (fMRI) data \cite{pang2023geometric}, both for task-related and resting brain states. The neuronal field theory model (NFT) is particularly noteworthy due to its simplicity, as it requires only one fixed and one adjustable parameter, in contrast to more complex wave propagation models. Contrary to the previous approach, it has been found that long-wavelength modes, specifically with wavelengths larger than 40 mm, play a dominant role in the propagation processes described by the geometric Laplacian, establishing a typical correlation length in the cerebral cortex. This finding extends beyond the traditional understanding of wave propagation in the cortex as a two-dimensional elastic medium described by an effective two-dimensional Laplacian and applies to three-dimensional samples involving non-cortical regions. Studies have shown that achieving a good fit of data requires fewer than 100 lowest modes of the 3D Laplacian \cite{pang2023geometric}. These findings suggest that the overall understanding of wave propagation in the brain needs to be reconsidered: instead of short-range correlations, long-range waves play a major role in brain activity.

Here, we employ spectral methods to investigate the impact of metric aspects on brain organization. We use two complementary approaches. First, we ask whether there is a generative model that incorporates intrinsic scale and can reproduce the essential spectral features of structural connectomes. A particular pattern of evolutionary laws has been proposed in \cite{pospelov2019spectral}, where it is argued that producing a true spectral density for organisms further up the evolutionary ladder requires a larger number of local conservation laws, but this model lacks any consideration of scale. Second, by examining the transition between localized and delocalized excitations propagating through a structural connectome, we investigate the presence of localized modes in the spectrum. This model implies the existence of a spatial scale that defines the localization length.

Here are some general remarks on the localization properties of the structural connectome. It would not be reasonable to expect all eigenmodes to be completely delocalized, as this would imply that the system is ergodic and it would be difficult to observe the effects of activity persistence. On the other hand, complete localization of the eigenmodes would result in the absence of correlations between brain regions. Therefore, the most likely scenario is that there is a coexistence of localized and delocalized modes. The initial discussion on this topic can be found in \cite{pospelov2019spectral}, although only a limited number of connectomes were analyzed in that study. Therefore, there is a need for further statistical verification of these claims and a more thorough analysis.


In \cite{pospelov2019spectral}, it was found that within the "soft" part of the Laplacian spectrum (at small eigenvalues corresponding to cluster structure), the level spacing distribution follows a deformed Poisson distribution. However, in the bulk of the continuous spectrum, it follows a combined Poisson-Wigner distribution. This behavior, from the perspective of Anderson localization in systems with diagonal disorder, suggests that the system is operating near the critical regime \cite{shklovskii1993statistics}. A mobility edge separates localized and delocalized modes, and although the structural connectome does not have on-site diagonal disorder like the conventional Anderson model, the inherent structural disorder plays a similar role. To determine the presence of criticality and mobility edge, it is necessary to analyze individual Laplacian and adjacency eigenmodes in more detail. One direct approach to address this issue is to examine the inverse participation ratio (IPR) of all eigenmodes. Elevated IPR values indicate localization, and we aim to elucidate the localization characteristics of these modes within the structural connectome. While we qualitatively support the findings of \cite{pospelov2019spectral}, which suggest the existence of localized modes in the bulk, we recognize that the interplay between these localized and delocalized modes is more complex.


The presence of localized modes in the structural connectome spectrum has also been mentioned in \cite{moretti2013griffiths}. In \cite{moretti2013griffiths}, it was conjectured that the connectome operates not at criticality, but rather in a Griffiths phase. This is more plausible from an operational perspective, as it does not require fine-tuning of parameters, in contrast to a purely critical situation. The Griffiths phase occurs when marks of criticality, such as power-laws, occur not only at a single value of a control parameter, but also within a range. In the conventional Anderson model, the diagonal disorder plays the role of a control parameter. However, in this work, only the case of structural disorder is considered; in this case, the Griffiths phase was found in \cite{munoz2010griffiths,odor2015griffiths}.

In the context of the Anderson localization the  candidates for the relevant rare states are the isolated modes at the edge of the spectrum corresponding to the clusters in the network \cite{nadakuditi2012graph}.
Later, it was also conjectured \cite{buendia2022broad} that the second type of critical point expected for the brain, the synchronization phase transition, could also be blown up into a Griffiths phase.
Different generic aspects of brain criticality can be found in reviews \cite{o2022critical,hesse2014self,beggs2022cortex,grosu2023fractal,cocchi2017criticality}.

In this study we turn to another spectral characteristic of the structural connectome not discussed in \cite{pospelov2019spectral,moretti2013griffiths} - the multifractality of eigenmodes, see \cite{evers2008anderson} for review. 
Previously, it was thought that multifractality was an intrinsic feature of the critical point. However, more recently, it has been recognized that it can also be a feature of the extended phase. This is usually referred to as the non-ergodic extended (NEE) phase or the extended multifractal phase. The new NEE phase has been first uncovered in the Generalized Rozentzveig-Porter model \cite{kravtsov2015random} and later was recognized in the clear-cut manner in the several models with matrix Hamiltonians \cite{kravtsov2015random, bogomolny2018eigenfunction, biroli2020anomalous, monthus2011anderson,motamarri2022localization,PhysRevB.105.174201,facoetti2016non,monthus2017multifractality}.

We are interested in the existence of multifractal modes in complex networks, and there are indeed relevant examples. Multifractality has been identified in the Anderson model with diagonal disorder on a Cayley tree \cite{sonner2017multifractality,monthus2011anderson,biroli2020anomalous}. It has been found that the fractal dimension of these modes depends on their position on the tree \cite{sonner2017multifractality}. Recently, an extended non-ergodic phase with multifractal eigenmodes was clearly identified in an Erdős-Rényi ensemble of graphs in the sparse regime with $p>1$ above the percolation threshold \cite{cugliandolo2024multifractal} without a diagonal disorder. It was argued that the mechanism behind multifractality may involve weakly interacting clusters.

In this paper, we will focus on the following issues:

\emph{A Novel Generative Model Incorporating Spatial Intrinsic Scale} We explore the principles behind the construction of a graph that resembles a connectome and could yield experimentally verifiable spectral density. For structural connectomes, we propose a generative model that employs nonlinear preferential attachment and an exponential parameter-dependent cutoff in edge lengths. This approach reproduces the geometric length distribution of fiber lengths found in the connectome as well as the spectral features and clustering properties.

\emph{Inverse Participation Ratio $(\text{IPR}_2)$ and localization of eigenfunctions}. We investigate the $\text{IPR}_2$ which indicates the localization properties of the Laplacian eigenmodes. Large values of IPR for some eigenmodes imply that these modes are localized in certain graph regions, while small values indicate delocalized states. Generically there are two types of the coexistence of localized and delocalized states: the existence of a mobility edge separating these states, and the presence of the ''scars'' corresponding to isolated localized states within the delocalized part of the spectrum. Our findings are as follows: there are states with high IPR values in the low-energy region of the spectrum that are localized at clusters, in agreement with \cite{pospelov2019spectral}. In addition, there are localized high-energy states with large IPR in the bulk that form two distinct small bands around the eigenvalues $\lambda=0$ and $\lambda=-1$ of the adjacency matrix. These eigenvalues correspond to connected and disconnected topologically equivalent nodes (TENs) similar to the pattern observed in \cite{kochergin2023anatomy}. Some of $\lambda=0$ modes are trivial TENs, which are the end points of the graph leaves. Therefore, despite the presence of localized states in the bulk, there is no sharp mobility edge in the bulk as suggested in \cite{pospelov2019spectral}. However, one could speak of an effective mobility edge between the bulk and the band of cluster isolated modes.

\emph{Multifractal Eigenmodes}. Multifractality is another potential localization-related statistical pattern that can be identified through the analysis of the $q$-dependent fractal dimensions $D_q$. These dimensions are defined by the variation of the inverse participation ratio $\text{IPR}_q$ over different scales. A value of $D_q = 0$ indicates localized modes, whereas $D_q = 1$ denotes extended modes. Modes with $0 < D_q < 1$ exhibit fractal behavior, and non-linearity in the variation of $D_q$ with $q$ reveals the presence of multifractality. We analyzed $D_q$ across various parts of the spectrum and unexpectedly found multiple modes with fractal dimension in the range $D_q= 0.7-0.8$, some of them with weak multifractality. These features of multifractality include the weak nonlinearity of the fractal dimension and the power-law scaling of return probabilities. With multiple reservations, we hypothesize that the structural connectome may operate in the multifractal regime and compare it with the Griffiths phase scenario.

\emph{Correlation of Clusters}. Soft modes exhibit a semi-Poisson distribution, which is consistent with the general arguments presented in \cite{avetisov2020localization}. The deviation of the level spacing distribution $P(s)$ from a pure Poisson distribution indicates interactions among cluster modes, similar to those extensively studied in \cite{kochergin2023anatomy}. It is noteworthy that recent spectral analysis of the C. Elegance structural connectome demonstrates the precise identification of clusters as the soft modes of Laplacian even in sparse scenarios \cite{onuchin2023communities}. To substantiate the presence of cluster interactions, we examine the correlations between the lowest eigenvalues $(\lambda_2, \lambda_3, \lambda_4)$. Given that the identification of $\lambda_2$ is straightforward -- it quantifies the number of links between hemispheres \cite{wang2017brain}—we investigate the behavior of $(\lambda_3, \lambda_4)$ as a function of $\lambda_2$. Our analysis reveals the clear fingerprint of interactions among the soft modes.

The paper is structured as follows. In Section 2, we present the generative model for the structural connectome. In Section 3, we compare various spectral characteristics of the model to experimental data. Section 4 analyses the eigenmode statistics and formulates a conjecture that the structural connectome operates in an extended non-ergodic multifractal phase. The findings are summarized in the Discussion.

\section{Description of a nonlinear geometric model}

Nowadays, the Watts-Strogatz (WS) model \cite{Watts1998} is considered a good "baseline" model that adequately describes many experimentally observed features of the brain's structural network. This model has a "small world" network structure, which combines short optimal paths with a high clustering coefficient. This interplay between these two properties is crucial for efficient brain function, providing significant advantages in signal processing. This structural organization is essential for healthy brain function. Indeed, deviations from these "small world" features have been observed in groups of patients with Alzheimer's disease, autism, and schizophrenia, as reported in \cite{Hilgetag2015}.

Since the pioneering work \cite{Friston2008}, more and more authors have claimed the ``hierarchical'' organization of the functional brain connectomes (see, for example \cite{hierarchy-mod, Akiki2019}). However, whether structural connectomes also have hierarchies consistent with functional ones is an open question that worries researchers around the world. \cite{hierarchy-struct-model, hierarchy-compl}.

Recent studies \cite{metric} have provided convincing evidence that the agreement between experimental data and mathematical models improves significantly when the metric structure of structural connectomes is taken into account. It has been shown that changes in brain network structure, in general, and changes in connection lengths, in particular, are linked to neurodegenerative diseases \cite{gollo2018fragility,dist_desease}.

Here we propose a simple model that has all the desired properties: (i) small-world behavior, (ii) a high clustering coefficient, and (iii) a spatial cutoff due to being embedded in a real three-dimensional space.

Specifically, we propose a model that combines the abstract generalized preferential attachment algorithm with an exponential penalty for long edges in a graph constructed from 3D coordinates of a human connectome. As it has been shown in \cite{krioukov}, the preferential attachment models with a scale-free node degree distribution (and hence with a high clustering coefficient) allow for a natural embedding into the hyperbolic Poincaré disc of finite radius. Due to their hyperbolic nature, scale-free networks are closer to trees than exponential Erdős-Rényi graphs. However, the scale-free networks created in this way are purely topological and do not contain any information about the spatial proximity of nodes in structural connectomes that exist in 3D space. The information about the metric structure of our model network was derived from the actual coordinates of nodes in human brain connectomes and was then applied to an artificial network that was created using the preferential attachment algorithm. This model is referred to as the Nonlinear Geometric Preferential Attachment (NGPA) model.

\subsection{Human connectome data}
We used the \href{https://braingraph.org}{braingraph.org database} \cite{PIT} for our analysis. This database contains structural connectomes of 426 human subjects  computed from high-angular resolution diffusion imaging data from the Human Connectome Project \cite{HCP}. Each connectome has been constructed at five different resolutions, and we used the graph of the highest resolution (with 1015 nodes) for further analysis, except for the investigation of multifractal properties. Each connectome was preprocessed in the following way: (i) isolated nodes and self-loops were removed; (ii) for graphs with more than one connected component, only the largest one was kept. Connectomes in which the largest component covered less than $80\%$ of the nodes were excluded from the dataset; (iii) edge weights were discarded and binary adjacency matrices were constructed. 

From each connectome in the dataset, we extracted the coordinates of all nodes for further construction of spatially embedded model networks. We ensured that the Euclidean distances between nodes connected by an edge correlate well with the actual length of the axonal fibers in the dataset (Pearson's correlation coefficient $r = 0.91$). Therefore, we verified the proxy relationship between the real and Euclidean distances \cite{Kaiser2011} and used the latter for our further analysis.

\subsection{NGPA algorithm}
Our network construction algorithm consists of the following steps:
\begin{itemize}
\item[-] Select and preprocess a specific human structural connectome with $N$ nodes and $E$ edges from the dataset of structural human connectomes \cite{PIT};
\item[-] Preserve the 3D coordinates of $(x,y,z)$ coordinates of all nodes and remove all existing links. Note that the set of 3D node coordinates is the only information used from the original graph to create its metric-aware counterpart. However, to test the impact of node positions on the model, experiments were conducted with randomly positioned nodes. The results were similar to those obtained with real node coordinates, suggesting that perturbations in node spatial embedding do not significantly affect the model's properties;
\item[-] Construct a new artificial network based on a set of nodes with known $(x, y, z)$ coordinates. We do this by using the generalized preferential attachment algorithm, which has two components: topological and metric. 
\begin{enumerate}
\item Abstract preferential attachment. Consider a scenario where some of the new artificial network has already been created. Let $j$ be a node within this existing part, which is connected to $d_j$ other nodes. Then, introduce a new node $i$ and connect it to node $j$ in the existing network with a probability that depends on the degree of node $d_j$, following the preferential attachment algorithm. The probability $P_{ij}$ of forming a link $ij$ is given by:
\be
P_{ij} = \frac{(d_j+1)^{\alpha}}{\sum_{k=1}(d_k+1)^{\alpha}}
\label{eq:pref-1}
\ee
where $\alpha\ge 0$ is the parameter of the model ($\alpha = 1$ corresponds to the standard ``rich gets richer'' linear preferential attachment model\cite{BAbook}). We have added 1 to $d_j$ in order to be able to define a non-zero connection probability for isolated vertices (i.e. when $d_j=0$). 
\item Spatial embedding. To construct a graph that inherits the structure of a real-world connectome embedded in 3D space, we penalize the formation of long links by introducing an exponential cutoff function, $e^{-r_{ij}/r_0}$. Here, $r_{ij}$ represents the spatial distance between nodes $i$ and $j$, which are constructed based on a dataset of 3D coordinates from a human connectome, and $r_0$ ($r_0>0$) is the cutoff parameter. We select $r_0$ as $r_0 = l_0/\beta$, where $l_0 = \la r_{ij} \ra = \sum_{i,j}r_{ij}/E \approx $ 15 mm is the average link length in the dataset.
\end{enumerate}
By combining the abstract preferential attachment algorithm with the spatial embedding, we define the new probability of link formation between nodes $i$ and $j$ in the three-dimensional network created by the generalized preferential attachment algorithm:
\be
p_{ij} = \frac{Q_{ij}}{\sum_{j}^N Q_{ij}}, 
Q_{ij} =  \frac{(d_j+1)^{\alpha}}{\sum_{k=1}(d_k+1)^{\alpha}}e^{-\beta r_{ij}/l_0}
\label{eq:pref-2}
\ee
\item[-] Each node creates $m$ connections with other nodes according to the probabilistic rule described above. No target node can be selected more than once (all connections must be unique). The number of new connections $m$ is randomly selected from the range $m \in [1, \frac{4E}{N}]$, which ensures that the network will have approximately the same density as the original graph. This network construction algorithm is applied separately to sets of nodes from the left and right hemispheres, and then $E_{interhem}$ inter-hemispheric connections are added to connect the two hemispheres of the model. These connections are created by randomly selecting a pair of nodes, one from each hemisphere. Although the number of inter-hemispheric connections $E_{interhem}$ is small compared to the total number of connections $E$, they are excluded from the analysis of edge lengths (only connections within each hemisphere are considered).
\end{itemize}

\subsection{Connection with other models}

For brevity, we will refer to the NGPA model as not only the result of the random network generation process described above, but also as the entire pseudo-connectome $M(\alpha, \beta)$ which consists of two model "hemispheres" (NGPA graphs) and their interhemispheric connections.

Note that standard network models can be derived through the two-parameter NGPA process with specific parameter values. For example, selecting $\alpha = 0$ and $\beta = 0$ ($M(0, 0)$) results in an Erdős-Rényi-like random graph \cite{er}. Selecting $\alpha=0$ and $\beta > 0$ imposes soft geometric constraints on network connectivity as in the random geometric graph model (but without a hard boundary). The limiting model $M(0, \infty)$ is approximately a $k$-nearest neighbors graph (since only the closest nodes are connected). On the other hand, NGPA models with preferential attachment and no geometric constraints ($\alpha > 0$ and $\beta = 0$) correspond to Barabási-Albert-like graphs, as described in \cite{nlBA}. These graphs have different regimes: $\alpha < 1$ corresponds to the subcritical regime, $\alpha = 1$ to the standard scale-free network, and $\alpha > 2$ to the "winner-takes-all" regime \cite{BAbook}.

\section{NGPA model versus real connectomes}

\subsection{Selecting optimal parameters}

\begin{figure}[ht]
\centering
{\includegraphics[width=16cm]{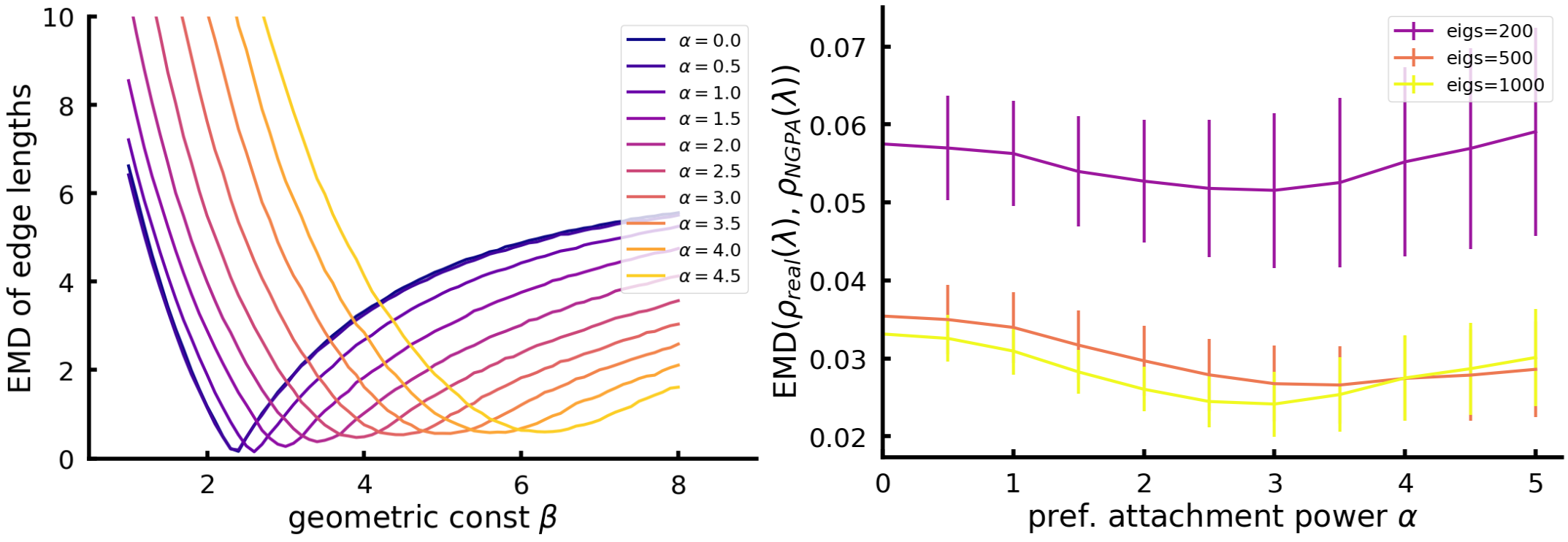}}
\caption{Left: EMDs between edge length distributions as a function of $\beta$. Each curve corresponds to a different $\alpha$ value. Right: EMD between the normalized Laplacian spectrum of connectomes and their corresponding NGPA models, as a function $\alpha$. The three curves are for distances calculated from the first 200, 500, and 1,000 eigenvalues, respectively. The curves are averages over 100 networks.}
\label{fig:params}
\end{figure}

To select the optimal parameters for the model, we followed a two-stage process. In the first stage, we determined the optimal geometric coefficient, $\beta_{\text{opt}}(\alpha)$, for each potential value of $\alpha$, where our search space was $\alpha \in [0, 5]$ and $\beta \in [0.8]$, see Fig. \ref{fig:params} for details. In the second stage, we selected the optimal value $\alpha_\text{opt}$ that minimized the "earth mover’s distance" (EMD) \cite{emd} between spectral densities of normalized Laplacian matrices from real connectomes and their NGPA equivalents, with corresponding optimal geometric constraints, $M(\alpha,\beta_\text{opt}(\alpha))$.

This procedure allowed us to reduce the computational complexity in order to avoid overfitting, while ensuring that the geometric structure of the optimal network coincides with that of the real connectome. In both cases, we used the Earth Mover's Distance as a measure of distance between distributions, as it takes into account metric space \cite{emd}. 

Both quantities exhibit a clear minimum, with a near-perfect match between edge length distributions at $\beta_{opt}$. For further analysis, we use an optimal NGPA model with $\alpha = 3$ and $\beta = 4.5$, which leads to a characteristic geometric scale of approximately 3.5 mm (see equation \ref{eq:pref-2}).

\subsection{Spectral density and Jaccard similarity}

Let us apply the NGPA model, as formulated above, to the real-world structural connectome data. To begin, we will focus on the simplest spectral characteristic - the spectral density, which can be defined for the graph Laplacian $L$ as follows:
\begin{equation}
 \rho(\lambda)=\frac{1}{N}\sum_i\delta(\lambda-\lambda_i)
\end{equation}
where $\lambda_i$ is the $i$-th eigenvalue of the Laplacian matrix $L$. Recall that the graph Laplacian is related to the adjacency matrix of the graph $A$ via the equation $L = D - A$, where $D$ is a diagonal matrix with the degrees of the nodes on the diagonal. The normalized Laplacian, $L^{\text{norm}} = D^{-1/2} LD^{-1/2}$, is often used to control the influence of hubs in collective dynamics, especially in heterogeneous networks such as connectomes. The normalized Laplacian has several advantages. Its spectrum is always in the interval $\lambda \in [0,2]$, so it can be used to compare networks with different sizes. Additionally, all eigenvalues of both $L$ and $L^\text{norm}$ are non-negative, with $\lambda_1 = 0$ and all other eigenvalues $\lambda_i > 0$. This property holds for connected graphs and is ensured by our preprocessing and network construction procedures.

\begin{figure}[ht]
\centering
{\includegraphics[width=16cm]{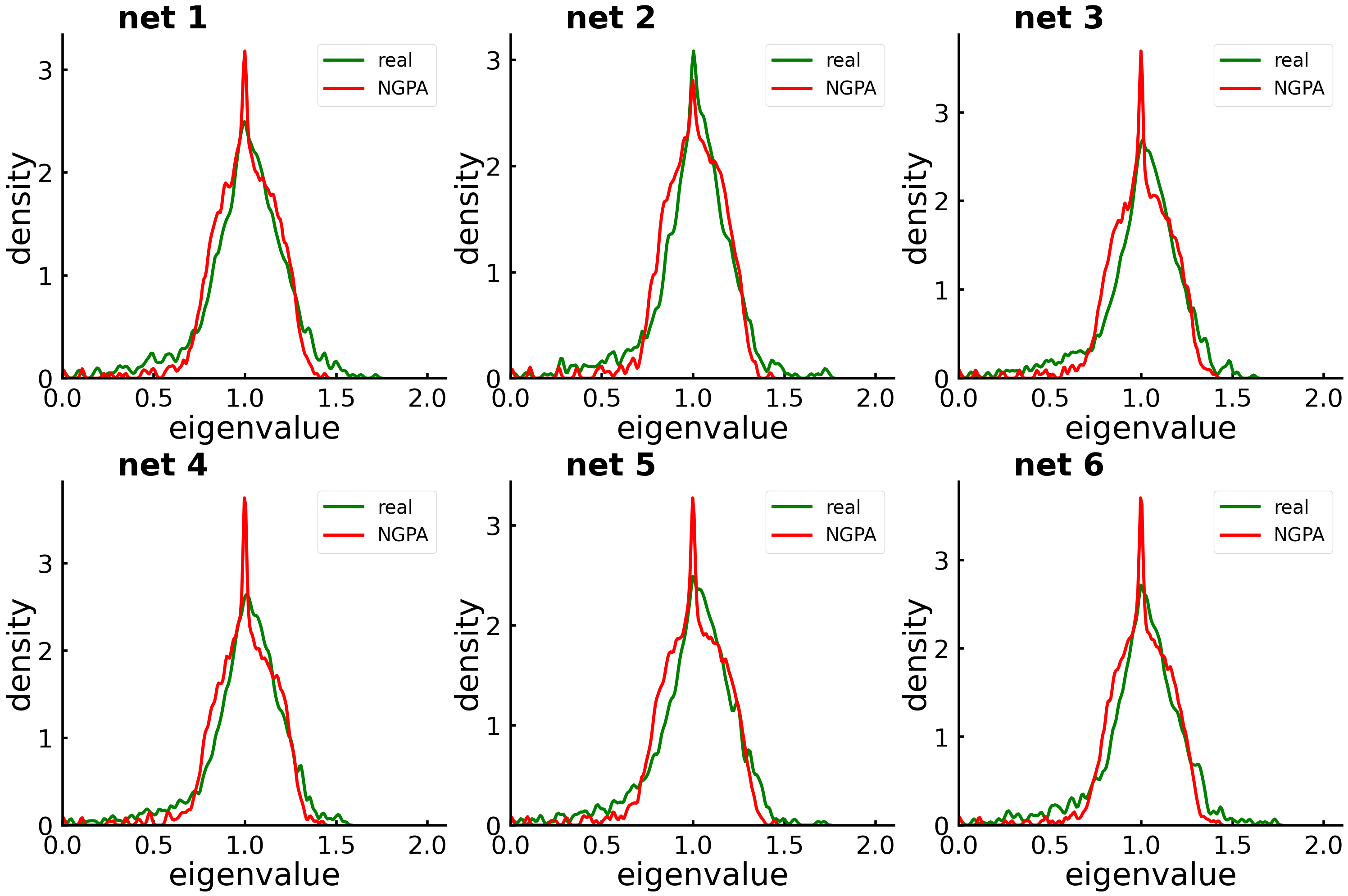}}
\caption{Spectral density of normalized Laplacian matrices for 6 real connectomes (green) and corresponding NGPA models (red). Model parameters $\alpha=3$ and $\beta=4.5$.}
\label{fig:NGPAspectrum}
\end{figure}

Typically, the Laplacian spectrum consists of several isolated soft (low-energy) modes corresponding to clusters, as described in \cite{nadakuditi2012graph}. These are accompanied by high-energy, continuous bulk modes. A comparison between the numerical data and the NGPA model in Fig. \ref{fig:NGPAspectrum} shows that the spectral density of the normalized Laplacian $L^{\text{norm}}$ in NGPA closely resembles the bulk spectral density calculated from experimental data. The number of clusters is correctly reproduced -- there are around 15 to 20 separated modes -- but their positions show slight shifts compared to those in the real connectome.

\begin{figure}[ht]
\centering
{\includegraphics[width=16cm]{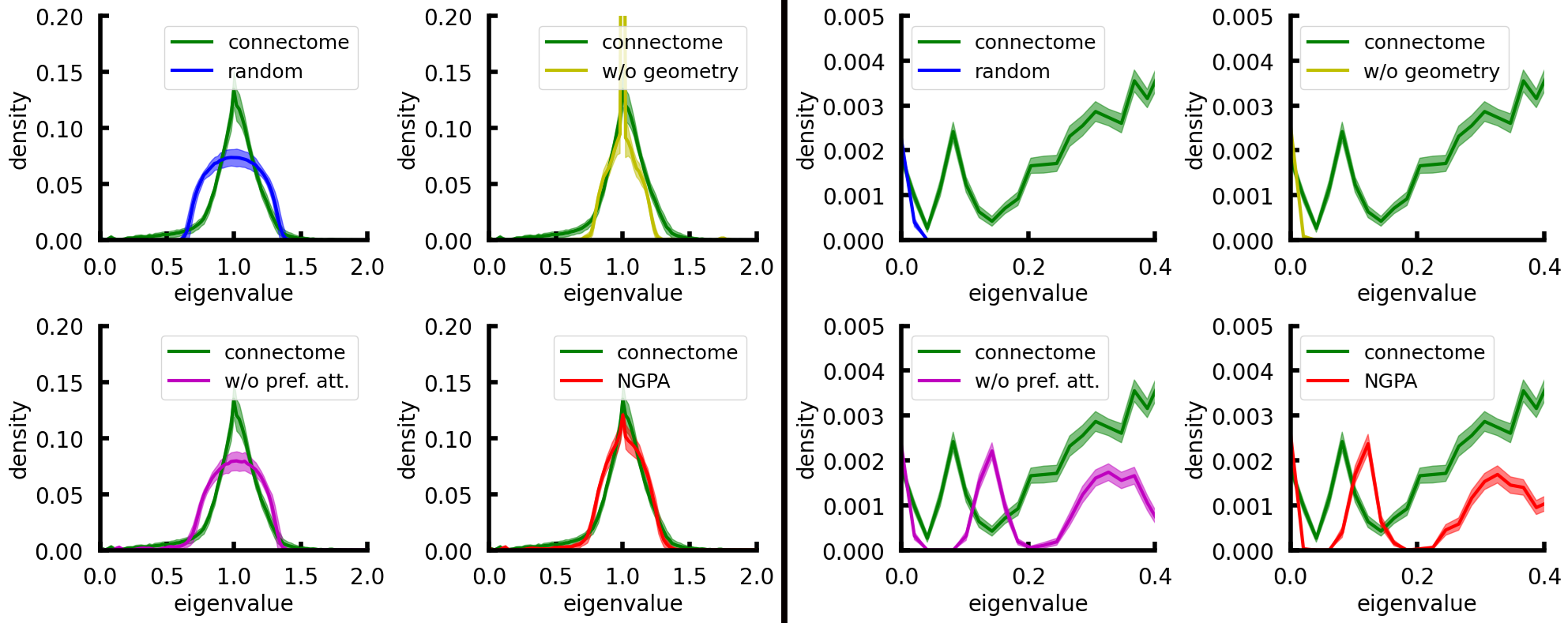}}
\caption{Spectral density of normalized Laplacian matrices for real connectomes and models. Left: full spectrum, right: insets for separated parts of the spectra. Models shown: "random" = $M(0, 0)$, "w/o geometry" = $M(\alpha_{opt}, 0)$, "w/o pref.att." = $M(0, \beta_{opt}(0))$, "NGPA" = $M(\alpha_{opt}, \beta_{opt})$. Spectra are averaged over 100 networks. Shaded regions indicate standard errors.}
\label{fig:NGPAspectrum_inset}
\end{figure}

It is worth noting that there is a strong peak in the spectral density of the normalized Laplacian in Fig.\ref{fig:NGPAspectrum_inset}. This peak in the Laplacian spectrum corresponds to a similar peak in the adjacency matrix spectrum at $\lambda = 0$. These peaks are indicative of a high level of heterogeneity in the graph \cite{silva2022analytic}. When a parameter is introduced that effectively measures heterogeneity, there is a transition at some point between spectra with and without peaks. Therefore, we can conclude that the connectome falls into the category of heterogeneous networks.

To test the performance of the NGPA model, we also compared the spectral features of the connectomes with simpler models that were obtained by removing some components from the original NGPA model $M(\alpha_{opt}, \beta_{opt})$. These models included: a random network $M(0,0)$, a model without geometric penalty $M(\alpha_{opt}, 0)$, and a model with no preferential attachment $M(0, \beta_{opt}(0))$. 

Differences in spectral density between the models and their comparison with connectome spectral density are shown in \ref{fig:NGPAspectrum_inset} (left). Note that both a random model and a geometry-only model have a semicircle-shaped spectral density, characteristic of simple random graphs \cite{ER_spectra} (more precisely, the limiting spectral density is the free convolution of a standard Gaussian distribution with Wigner’s semi-circular law \cite{Ding2010}, since we are interested in the spectrum of $L^{\text{norm}}$). The only significant difference between these two models is the presence of a low-energy eigenvalue region in the geometry-aware model, that corresponds to the natural formation of connected clusters under geometric constraints (fig. \fig{fig:NGPAspectrum_inset}, right). However, this zone is significantly shifted right compared with the real spectra, indicating weaker cluster structure. 

As for the model with preferential attachment but without geometric constraints ($M(\alpha_{opt}, 0)$), its spectral density has a more triangular shape due to the partially scale-free structure \cite{Farkas2011}. Together with the NGPA model, they provide a better approximation to the spectral density of the bulk than homogeneous models. This bulk shape is also typical for structural connectomes of other organisms, including cats and monkeys \cite{deLange2014}. However, one important difference id found in the low-energy region: while $M(\alpha_{opt}, 0)$ has no such eigenvalues (except for one separated eigenvalue $\lambda_2$, which is always present in two-hemispheric graphs), NGPA model exhibits a clear peak in low eigenvalue density whose position is close to the one in real connectomes.

Note that both the $M(\alpha_{opt}, \beta_{opt})$ (NGPA) model and the $M(\alpha_{opt}, 0)$ (geometry-free) model have a peak in the density of eigenvalues around $\lambda = 1$, as shown in \fig{fig:NGPAspectrum_inset}. This peak corresponds to localized states in the network, which arise from the presence of nodes with similar connectivity. This phenomenon has also been observed in real connectome spectra. We discuss this further in Section C. However, only the NGPA model accurately reproduces this peak, while the geometry-free model significantly overestimates it (its spectral density in \fig{fig:NGPAspectrum_inset} is truncated from above for better visibility). This is because the geometry-free approach can lead to the formation of super-hubs, which forcefully attract connections from weaker nodes.

\begin{figure}[ht]
\centering
\includegraphics[width=16cm]{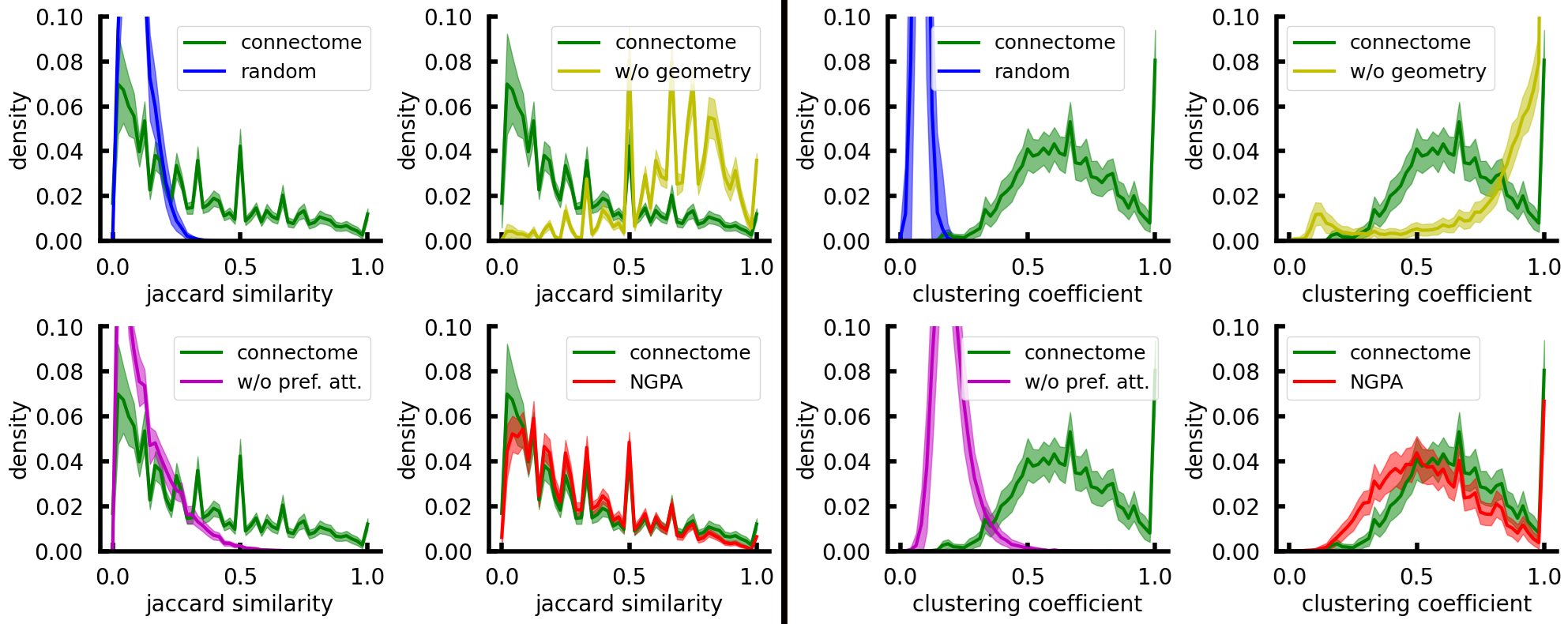}
\caption{Left: distributions of Jaccard similarity matrix elements in the real connectomes and the NGPA models. Right: the same for local clustering coefficients distributions. Curves are averaged over 100 networks, shaded regions indicate standard errors.}
\label{fig:NGPAsim}
\end{figure}

It is instructive to compare the NGPA model with the generative model proposed in \cite{pospelov2019spectral}, which reasonably well reproduces the spectral density as well. Both models highlight the significance of the local structure of the connectome. In the NGPA model, local properties are incorporated through the exponential cutoff of edges that significantly exceed $r_0$ in length, imposing a spatial constraint on the preferential attachment process. Conversely, the generative model proposed in \cite{pospelov2019spectral} addresses local properties  by introducing additional constraints associated with the local clustering that govern network evolution. The number of such additional constraints is proportional to network size $N$. In both cases, numerous local constraints suppress network growth. It is worth noting, however, that in the NGPA model this behavior is achieved by controlling a single tuning parameter $r_0$, which defines the geometric scale. Combined with the non-linear preferential attachment mechanism, it is sufficient to reproduce quantitatively the distribution of local clustering coefficients of nodes (\fig{fig:NGPAsim}). This means that high transitivity of brain's structural networks and even local clustering structure, which was important in \cite{pospelov2019spectral}, can be obtained as a by-product of the combination of the two structural principles outlined above.

The NGPA model also reproduces well the topological overlap matrix, $O_{ij}$ of the real connectome. Following \cite{Ravasz2002}, the matrix $O_{ij}$ is defined as follows: 
\begin{equation}
O_{ij}=\frac{J_N(i,j)}{\min(k_i,k_j)},
\end{equation}
where $J_N(i,j)$ is the number of common neighbors of node $i$ and $j$ (+1 if there is a direct link between $i$ and $j$), $k_i$ is the degree of node $i$. A topological overlap of 1 between nodes $i$ and $j$ implies that they are connected to the same vertices, whereas a 0 value indicates that $i$ and $j$ do not share links to common nodes among the nearest neighbors. The corresponding distributions of $O_{ij}$ elements is shown in \ref{fig:NGPAsim}, left. Note that the peaks of this distribution correspond to a high number of linked node pairs with disjoint common neighbor sets (or, alternatively, node pairs having a single common neighbor): $O_{ij} = \frac{1}{2}, \frac{1}{3}, \frac{1}{5}$, etc. The NGPA model correctly reproduces these peaks, at the same time slightly overestimating their amplitude. We attribute this to the preferential attachment mechanism acting on the scales smaller than $r_0$ and thus not being suppressed by the geometric penalty, which stimulates formation of "star-shaped" subgraphs.

\subsection{Hyperbolic embedding}

Hyperbolic random graphs are a promising category of geometric graphs that share many characteristics with complex real-world networks, including a power-law degree distribution, small diameter and average distance, and a high clustering coefficient \cite{krioukov}. Important open questions include determining the embedding of a real network in a hyperbolic space and identifying the signature of a hyperbolic manifold from network data. 

One may question why the hyperbolicity of the brain gains the interest of researchers and what insights can be gained from measuring hyperbolicity. It is worth noting that hyperbolic manifolds cannot be embedded in any Euclidean space with a fixed dimension. However, they can be embedded in a space where volume grows exponentially with the radius. These spaces are called "hyperbolic". The more "hyperbolic" the manifold, the greater the space it explores, but the brain network is constrained by its confinement in the finite three-dimensional cranium.

\begin{figure}[ht]
\centering
{\includegraphics[width=16cm]{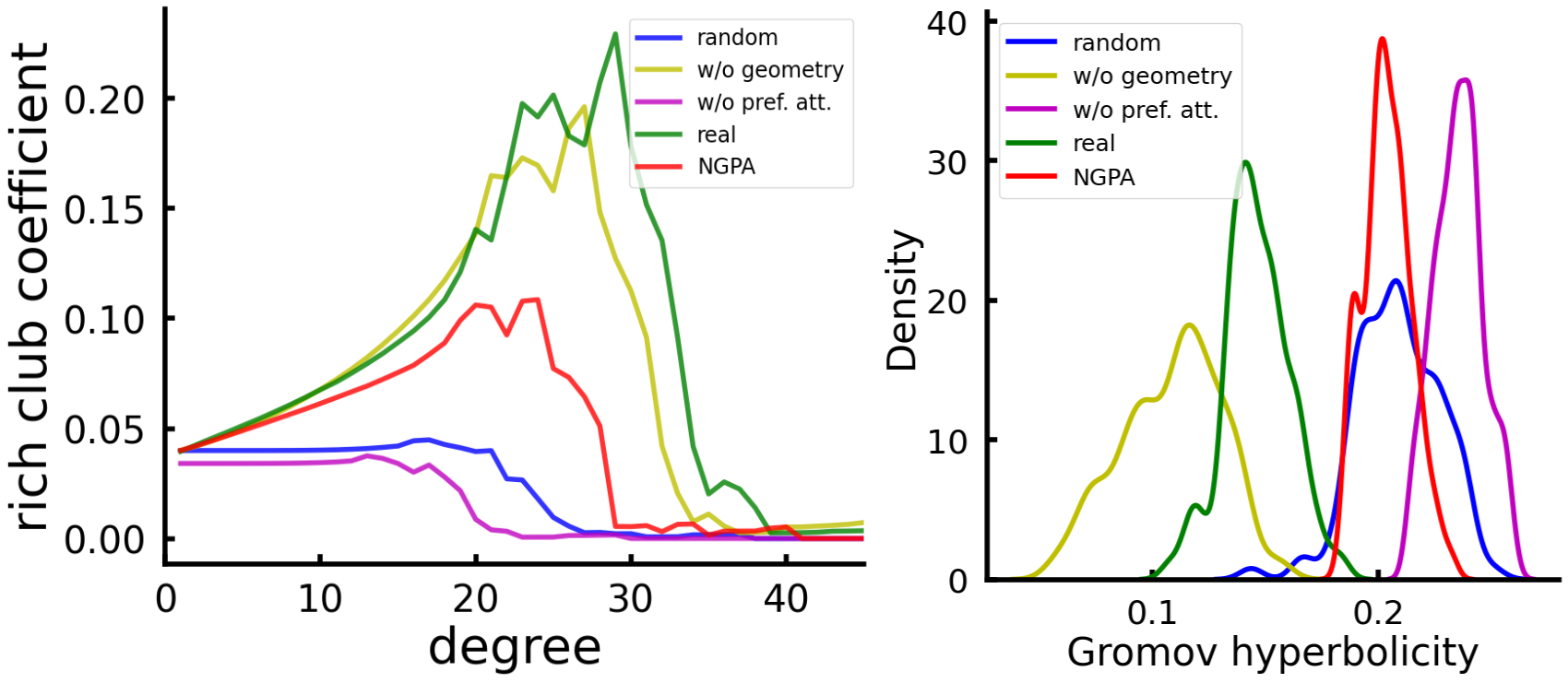}}
\caption{Left: rich-club coefficients for real connectomes and model networks. Right: the same for Gromov hyperbolicity indices distributions. Models shown: "random" = $M(0, 0)$, "w/o geometry" = $M(\alpha_{opt}, 0)$, "w/o pref.att." = $M(0, \beta_{opt}(0))$, "NGPA" = $M(\alpha_{opt}, \beta_{opt})$. The distributions are averaged over 100 networks.}
\label{fig:rc-hyp}
\end{figure}

The conflict between the exponential growth of the network and the limited volume it occupies inevitably leads to the formation of "crumples", resulting in the emergence of "shortcuts" in the embedded manifold. These shortcuts may create additional connections that reduce the time required for signals to travel from one brain area to another. In other words, one can hypothesize that a more hyperbolic brain operates faster. It's interesting to note that the cutoff parameter in the NGPA model, obtained through an optimization process as $r_0 = l_0 / \beta = 3.5 mm$ (see Eq. (\ref{eq:pref-2})) correlates with the typical cortical thickness, which is about 2.5 mm \cite{Tallinen2016}.


The hyperbolicity of a graph was defined by Gromov \cite{Gromov1987} (we are only discussing here the so-called 4-points condition). Let $v_1$,$v_2$,$v_3$,$v_4$ be the vertices of the graph and let $S_1$,$S_2$ and $S_3$ be defined as follows:
\begin{equation}
\begin{array}{l}
S_1 = d(v_1,v_2)+d(v_3,v_4) \medskip \\
S_2= d(v_1,v_3)+d(v_2,v_4) \medskip \\
S_3=d(v_1,v_4)+d(v_2,v_3)
\end{array}
\label{eq:hyp01}
\end{equation}
where $d(v_i,v_j)$ is the shortest path length between $v_i$ and $v_j$. Take $M_1$ and $M_2$ -- the two largest values among $S_1$,$S_2$ and $S_3$. We define the hyperbolicity for node set $(v_1,v_2,v_3,v_4)$ as 
\begin{equation}
\delta(v_1,v_2,v_3,v_4)=M_1-M_2,
\label{eq:hyp02}
\end{equation}
The hyperbolicity $\delta(G)$ of the entire graph is the average over all possible 4-points hyperbolicities:
\begin{equation}
\delta(G)=\langle \delta(v_1,v_2,v_3,v_4)\rangle_{v_1,v_2,v_3,v_4 \in V(G)}
\label{eq:hyp03}
\end{equation}

Using the definition \eqref{eq:hyp03}, we calculated the Gromov hyperbolicity for connectomes and model networks. The results are shown in Fig. \ref{fig:rc-hyp} (lower values of $\delta(G)$ correspond to ``larger hyperbolicity''). We found that structural connectomes exhibit pronounced hyperbolic properties. Model $M(\alpha_{opt}, 0)$ had the lowest hyperbolicity index, which is due to its highly heterogeneous structure and dependence on a small set of superhubs that other nodes connect to. Other models, including NGPA, did not replicate the observed hyperbolic behavior of the connectome, though the NGPA model came closest. We attribute this to the rich-club organization in human connectomes (Fig.\ref{fig:rc-hyp}, left), which the NGPA lacks (see Discussion).

The rich club coefficient is defined as 
\be
\phi(k) = \frac{2E_{>k}}{N_{>k}(N_{>k} - 1)}
\label{eq:rc}
\ee

It measures the "network core" density, which is known to be high in brain networks \cite{richclub} and claimed to be important for efficient information processing.

\section{Statistics  of eigenmodes}

\subsection{Inverse Participation Ratio $\text{IPR}_2$ across the spectrum}. 

Let us turn to a deeper study of the localization properties of eigenmodes. Traditional indicators for localization or delocalization of eigenmodes include: (i) the level spacing distribution $P(s)$, (ii) the $r$-statistics for the real spectrum and (iii) the inverse participation ratio $\mathrm{IPR}$. In our previous work \cite{pospelov2019spectral}, we exclusively focused on $P(s)$ and discovered a hybrid Wigner-Poisson distribution in the continuous region, alongside a deformed Poisson distribution in the clustered segment of the spectrum.

The level spacing distribution, $P(s)$, and $r$-statistics serve as integral descriptors of the spectrum. Therefore, to investigate the localization properties of individual modes, it is useful to focus on the inverse participation ratio $\mathrm{IPR}$. In this discussion, we direct our attention to the $\mathrm{IPR}$ of individual modes, large $\mathrm{IPR}$ values signify localized states. Subsequently, we denote by $\mathrm{IPR}_{q}^{(i)}$, the inverse participation ratio for the eigenvector $\psi_i(n)$ where $i$ counts the eigenvectors and $n$ denotes the $i^{\rm th}$ component of the eigenvector $\psi_i(n)=
\{\psi_i(1),\psi_i(2),...,\psi_i(N)\}$:
\begin{equation}
\mathrm{IPR}_{q}^{(i)}=\sum_{n=1}^N \left|\psi_i(n)\right|^{2q}
\label{eq:IPR(E)}
\end{equation}

\begin{figure}
\centering
\includegraphics[width=16cm]{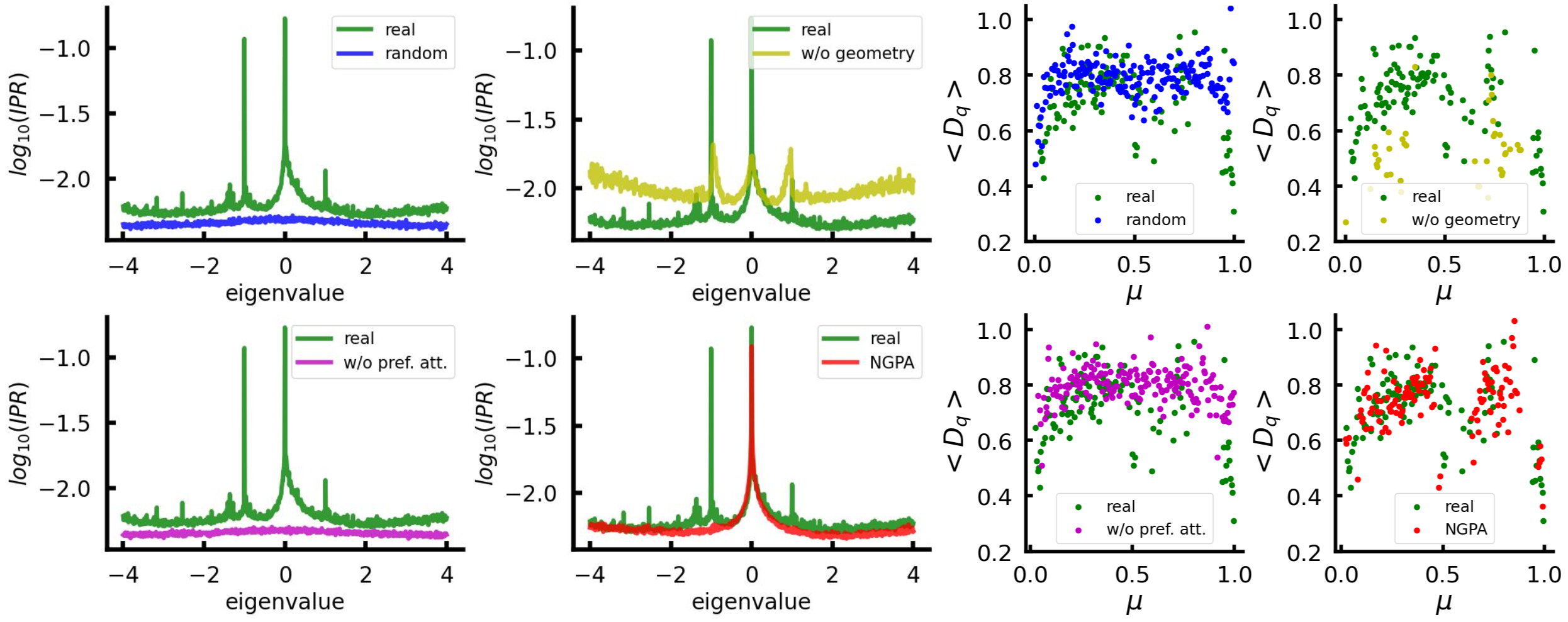}
\caption{Left: inverse participation ratios ($\text{IPR}_2$) of the adjacency matrices' eigenvectors as functions of corresponding eigenvalues. Right: average fractal dimensions of eigenmodes with a certain position $\mu$ on the spectrum ($\mu$ is defined as $\mu_i(\lambda_i) = \frac{i}{N} , i \in [1,N]$, see Appendix \ref{appendix:mfr} for details of multifractal eigenmodes detection). Data are shown for connectomes and network models (averaged over 100 networks). Models shown: "random" = $M(0, 0)$, "w/o geometry" = $M(\alpha_{opt}, 0)$, "w/o pref.att." = $M(0, \beta_{opt}(0))$, "NGPA" = $M(\alpha_{opt}, \beta_{opt})$.}
\label{fig:IPR}
\end{figure}

Mixed statistics as reported in \cite{pospelov2019spectral} indicates the presence of localized states within the bulk of the spectrum. Analyzing the $\text{IPR}_2$, we did identify such states. However, their nature is distinct: they exhibit significant peaks at $\lambda=0$ and $\lambda=-1$ in the spectrum of adjacency matrix as depicted in Fig.\ref{fig:IPR}. Notably, the positions of these peaks are not arbitrary  they correspond to scar-like localized states, which have been extensively studied in \cite{kochergin2023anatomy}. The existence of these states has been revealed within networks containing topologically equivalent (TEN) nodes, which possess similar connectivity to their surroundings. A connected (interacting) pair of TENs gives rise to a peak at $\lambda = -1$, whereas a disconnected pair manifests at $\lambda = 0$. In cases where TENs are non-ideal, a distribution of large IPR values around $\lambda=\{0,1\}$ emerges, as is explained in \cite{kochergin2023anatomy}.
Some of the $\lambda=0$ states are trivial TEN's localized at the ends of the leaves. Such states have been previously discussed for heterogeneous networks \cite{Tapias_2023,silva2022analytic}. However, some of $\lambda=0$ states are localized at a small amount of nodes in the bulk of a connectome.

Fig. \ref{fig:IPR} illustrates perfect identification of the $\lambda=0$ peak, while the $\lambda=-1$ peak is absent in the NGPA. The explanation for this discrepancy is as follows: the $\lambda = -1$ peak corresponds to interacting TENs at some distance. If the cut-off scale introduced is smaller than the typical distance between interacting TENs, they are artificially suppressed, leading to the absence of the $\lambda=-1$ peaks. To test whether this explanation is valid, we looked for emergence of $\lambda=-1$ peaks by increasing the cutoff scale. As expected, these peaks appeared when the cutoff scale was increased by a factor of 2.5-3, demonstrating that the typical scales responsible for averaged spectral density and correlations between individual TENs differ slightly. As anticipated from a broader perspective, we also observe states with significant IPR in cluster states \cite{avetisov2020localization,kochergin2023anatomy}. Eigenmodes associated with cluster modes exhibit localization across multiple clusters.

\subsection{Multifractality of connectome eigenmodes}.

Higher $\text{IPR}_q$ provide the possibility to investigate the fractality and multifractality of the spectrum.
The multifractality can usually be analyzed via two related variables: the fractal dimension, $D_q$, and the fractality spectrum, $f(\alpha)$, \cite{evers2008anderson}. Consider the size dependence of $\text{IPR}_q$
\begin{equation}
\mathrm{\text{IPR}}_{q}\sim N^{-\tau_q}, \quad \tau_q=D_q(q-1),
\label{eq:Dq}
\end{equation}
where the fractal dimension for a given $q$ is defined in the interval $0<D_q<1$. For localized states $D_q=0$, and for completely delocalized states $D_q=1$. The multifractality spectrum $f(\alpha)$ is set by the Legendre transform of the fractal dimension

\begin{equation}
\tau_q= min_{\alpha} (q\alpha -f(\alpha))
\end{equation}

If $D_q$ has non-trivial $q$-dependence, the state is multifractal in a strict sense. Otherwise, we shall refer to it as a fractal state. We can consider the fractal dimension of individual states as well as the averaged fractal dimension over the spectrum, here we focus on the former option. In the case of weak multifractality, the fractal dimensions can usually be presented in the form
\be
    \tau_q= d(q-1) +\gamma q(1-q), \qquad \gamma<<1
\label{eq:quadr}
\ee

We investigated the fractal dimensions of eigenmodes in real structural human connectomes and random models. The advantage of the dataset we used was that the structural connectomes were built at different scales, with each one approximately twice the size of the previous one (60, 125, 250, 500 and 1000 nodes). We excluded the smallest networks with a size of 60 nodes from our analysis of multifractal properties, leaving us with four networks at four different spatial scales to study the dynamics of eigenstates as the system size changed.

\begin{figure}
\centering
\includegraphics[width=16cm]{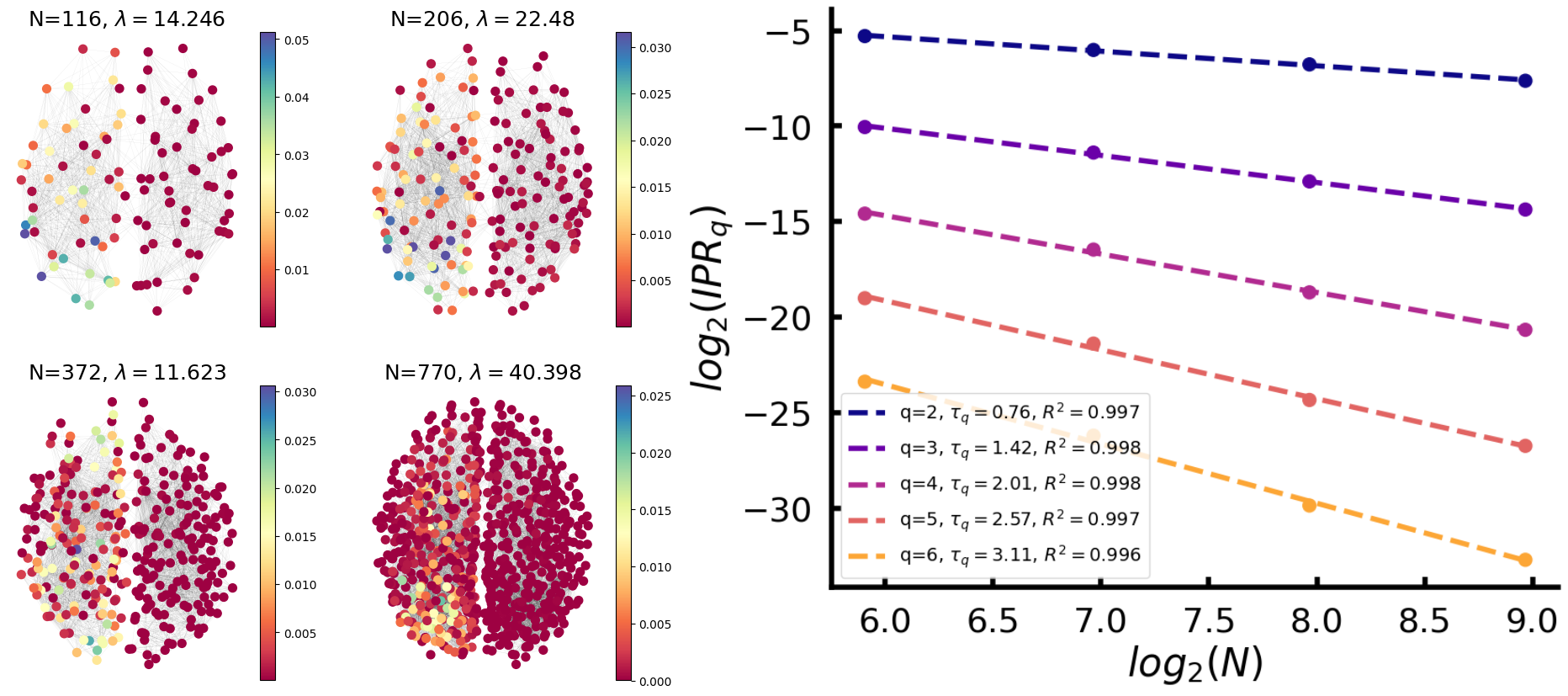}
\caption{Left: visualization of a single eigenmode across 4 different resolution scales. The color encodes the squared eigenvector elements, which represent the probability distribution over nodes. Right: $\text{IPR}_q(N)$ dependence for different $q$ values, along with the corresponding linear fit. The values of $\tau_q$ are shown in the legend.}
\label{fig:IPRq}
\end{figure}

\begin{figure}[ht]
\centering
{\includegraphics[width=16cm]{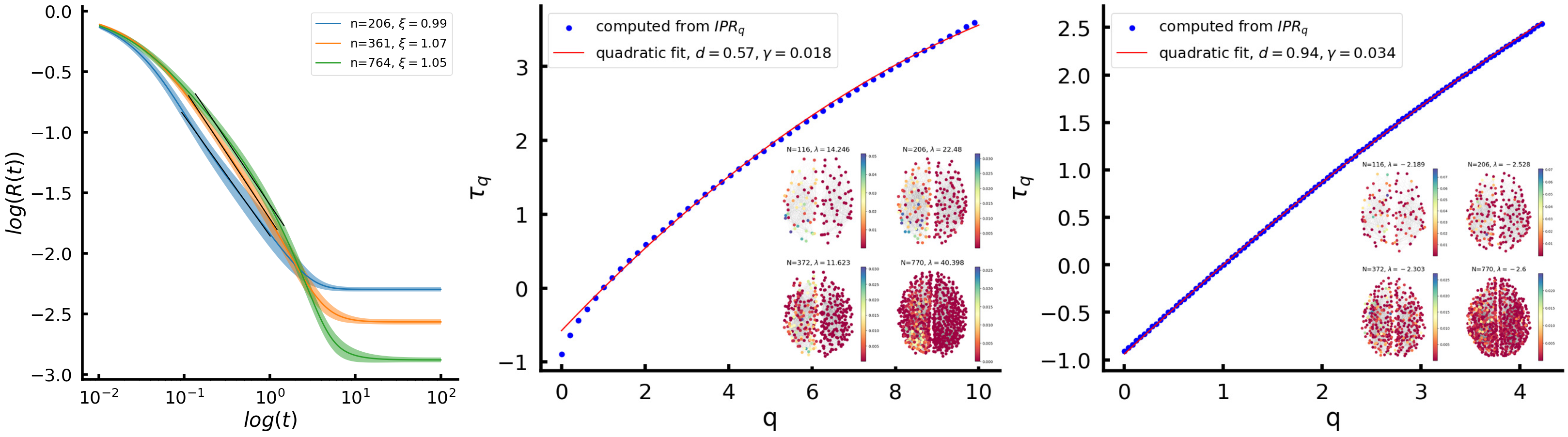}}
\caption{Left: power-law region in the return probability $R(t)$ across 3 different scales. Curves are averaged over 100 networks, for all scales we find $\xi \approx 1$. Central: $\tau_q(q)$ dependence for a single out-of-bulk eigenmode localized in the left hemisphere. Right: the same for an in-bulk eigenmode.}
\label{fig:mfr}
\end{figure}

First, we found multiple fractal states in the spectrum with a clear-cut linear dependence for $\tau(q)$, see Fig.\ref{fig:IPRq}. These states are mostly distributed in the bulk of the spectrum and have a non-trivial distribution of the fractal dimensions $D_q(\lambda)$. Next, we looked for deviations from linearity in $\tau (q)$ to confirm multifractality. Assuming a quadratic fit \eq{eq:quadr} we have found multiple multifractal states with small $\gamma$ and parameter $d$ depending on the position of the eigenmode in the spectrum. Fig \ref{fig:mfr} shows examples of typical eigenmodes along with their corresponding $\tau_q(q)$ plots. The quadratic fit confirmed that the system was in a weak multifractal regime ($\gamma<<1$) and also showed lower values of $d$ for eigenmodes modes outside the bulk corresponding to clusters (see also Fig. \ref{fig:IPR} right).

One more indication of a multifractal network state 
is the power-law in the return probability $R(t)$ defined as 
\be
R(t)= \frac{1}{N}\sum_i{e^{-t\lambda_i^\text{lap}}} \qquad R(t)\sim t^{-\xi}
\label{eq:r(t)}
\ee
where $\lambda_i^\text{lap}$ is the $i$-th eigenvalue of the network Laplacian $L$. We have found the power-law scaling for $R(t)$ with $\xi \approx 1$, see \fig{fig:mfr} which suggests that the decay of eigenstates is slower than it would be in a fully delocalized system.

\subsection{Correlation of cluster modes}

\begin{figure}[ht]
\centering
\includegraphics[width=16cm]{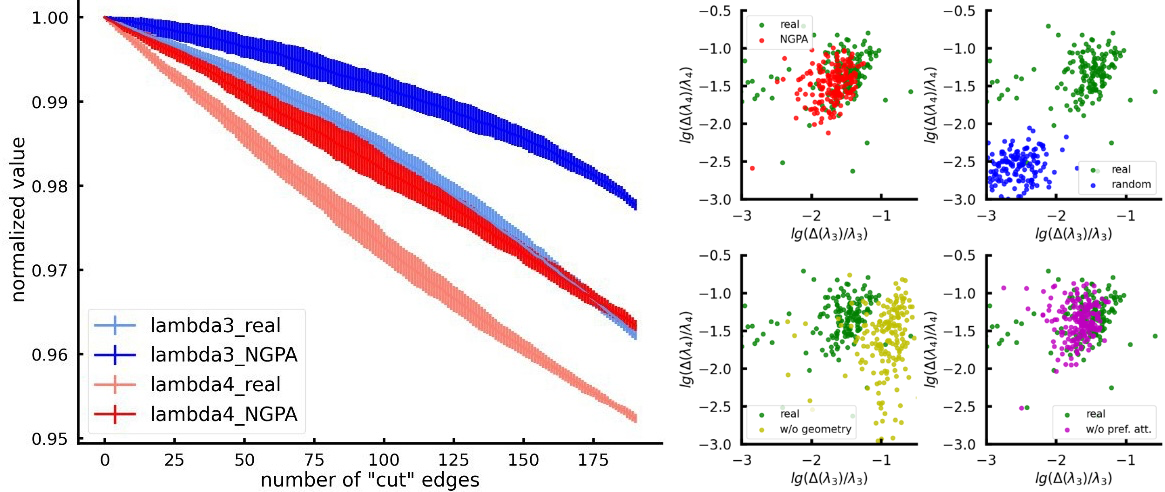}
\caption{Left: example of $\lambda_3$, $\lambda_4$ dynamics as interhemisheric edges are progressively removed from the connectome and from its NGPA model. The bars indicate standard deviations calculated from 100 random orderings of interhemispheric edges cuts. Right: relative changes of $\lambda_3$ and $\lambda_4$ after removing all interhemispheric edges from connectomes and model networks. Each dot represents a single network. Models shown: "random" = $M(0, 0)$, "w/o geometry" = $M(\alpha_{opt}, 0)$, "w/o pref.att." = $M(0, \beta_{opt}(0))$, "NGPA" = $M(\alpha_{opt}, \beta_{opt})$.}
\label{fig:lambdas}
\end{figure}

Our results on multifractality indicate that the community structure of connectomes may lie in the realm of interacting clusters as discussed, for example, in \cite{kochergin2023anatomy}. To validate the presence of correlations among the soft modes of the normalized Laplacian $L^\text{norm}$, we study the behavior of its ordered eigenvalues $\lambda_3$ and $\lambda_4$ as $\lambda_2$ varies. Recall that $\lambda_2$ is linearly proportional to the minimum number of links necessary to partition the network into two disjoint components \cite{chung1997spectral}. In our context, $\lambda_2$ approximately counts the number of connections between the two hemispheres of the brain \cite{wang2017brain}. As expected, the number of clusters correlates with the number of isolated soft modes of Laplacian. To test their independence, we sequentially cut all interhemisperic edges in connectomes and random models and investigate the dynamics of $\lambda_3$ and $\lambda_4$ (see Fig. \ref{fig:lambdas}).

We observe that the $M(0,0)$ random  model fails to accurate reproduce the correlation among soft modes, with changes in $\lambda_3$ and $\lambda_4$ being $\approx 10$ times weaker than in real networks. At the same time, both geometry-aware models quantitatively reproduce the effect (see Fig. \ref{fig:lambdas} right). The explanation for the symmetry-breaking behaviour of the $M(\alpha_{opt}, 0)$ model as well as geometric distribution of the first normalized Laplacian eigenvectors, are given in Appendix \ref{appendix:vis}.

Thus, we conclude that anomalous coupling between cluster modes (as compared to a random network) does really exist. We also find that this coupling is purely geometric, arising from the limited edge lengths and resulting local hemispherical structure.

\subsection{Do structural connectomes operate in the Griffiths phase or in the extended multifractal phase?}

The clear-cut fractal nature of eigenmodes and the hallmarks of multifractality lead us back to the question of the criticality of the structural connectome, specifically the relation to the Anderson localization of eigenmodes. There are two possible scenarios: (i) a critical point at a fixed value of a control parameter or (ii) a phase over a range of control parameters. The latter option is  more desirable due to its lack of fine tuning, which seems much more likely from a biological standpoint. Two options host the stretched Anderson critical point: the Griffiths phase and the extended non-ergodic phase (NEE phase). These are believed to be distinct, despite having some similarities.


In both cases it is expected that heterogeneity is the source of the critical phase, however the selection of the proper measure of heterogeneity as the control parameter is not evident. In the Griffiths phase the key players are the rare effects which correspond to spectral edges, similar to Lifshitz tails or isolated cluster eigenvalues for the graph Laplacian. The power-law behavior is found in some interval of the control parameter. The NEE phase has a typical wavefunction with a few maxima located at several points in the whole system, manifesting multifractality, and the fractal dimension $D_q$ is $q$-dependent. A possible mechanism for the NEE phase's existence is the presence of minibands in the bulk \cite{khaymovich2020fragile, altshuler2023random}. At large $N$, the number of states in a miniband diverges while its width decreases, resembling our spectra with a peak at $\lambda = 0$ in the spectral density (see Fig.\ref{fig:NGPAspectrum_inset}). The origin of the NEE phase within the framework of the renormalization group was discussed in \cite{zirnbauer2023wegner} and \cite{altshuler2024renormalization}.

Recently, an example of a system with a multifractal non-ergodic phase with a clear control parameter was presented in \cite{cugliandolo2024multifractal}. This was found in the Anderson model on a weighted Erdős-Rényi ensemble of random graphs with no diagonal disorder in the sparse regime above the percolation threshold $p = 1$. The single-particle system on the random graph exhibits the NEE phase in a certain range of the $(\lambda, p)$ plane. Several variables calculated in the model support the correctness of this phase identification. The entire spectrum of the model includes separated modes corresponding to clusters and bulk. There are several degenerate $\lambda = 0$ bulk states in the adjacency matrix spectrum, which are argued to be localized. These states disappear at high $p$. Cluster modes are also localized.

The authors of \cite{cugliandolo2024multifractal} propose a peculiar mechanism responsible for the emergence of the multifractal phase. They argue that weakly interacting clusters play a role. The number of clusters is small and the number of connections between them is also small. It has also been argued that the presence of local modes at $\lambda = 0$ does not significantly affect multifractality. Note that the $q$ dependence of the fractal dimension is not quadratic in this model, so this finding provides an example of an extended phase with properties similar to both Griffiths and NEE phases.

Turning back to our study, let us note that we do not have any tool to analyze connectomes analytically, but the spectral properties we found numerically are suggestive. Indeed, we have a number of separated localized cluster-related modes as well a peak at $\lambda = 0$ in the bulk, populated by localized modes with a high $\text{IPR}_2$. Additionally, there is a clear-cut IPR peak at $\lambda = -1$. The cluster modes interact with each other, yielding non-Poisson statistics in the cluster band (see Fig. \ref{fig:lambdas}). Therefore, these findings are similar to those reported in \cite{cugliandolo2024multifractal}, and indeed, the fractality of the multiple modes is identified. At least half of the extended modes are fractal with a typical fractal dimension $D_2=0.7-0.8$ (see Fig. \ref{fig:IPR} right, note that both modes outside the bulk and those around $\lambda = 0$ have significantly lower $D_q$). More refined fitting for $D_q$ reveals the weak multifractality, although the deviation from the linearity is small. We also find power-law behavior in the return probability, analogously to \cite{cugliandolo2024multifractal}. Of course, the architectures of the connectome and the model in \cite{cugliandolo2024multifractal} differ since we are not dealing with a sparse case.

Therefore, to make any firm statement concerning the relevance of the NEE phase for the structural connectome, we need to introduce a proper control parameter that measures heterogeneity, like the one considered in \cite{silva2022analytic}, and identify a possible critical interval for that control parameter. This point definitely requires detailed study. There are many reservations about the idea that the connectome operates in a nonergodic extended phase. For one thing, the size of the system is small, and we cannot rule out the possibility that weak multifractality in the spectrum is a finite-size effect. Also, the analysis only involves four points at different $N$, and the directions of connectome links are not considered. Nevertheless, despite all these subtleties, we suggest that the connectome does operate in NEE phase, although it might be very close to the Griffiths phase within its general framework.
 
\section{Discussion}

It is known that the brain seeks to find the optimal balance between the efficiency of information processing, which requires a large number of connections, and the efficient use of metabolic resources needed to maintain these connections \cite{bullmore2012economy}. Various topological features of brain networks, such as the small-world structure and the presence of highly connected hubs, are consequences of the need to maintain this equilibrium. In this study, we developed a simple two-parameter model of structural connectomes that incorporates both of these aspects and is able to replicate many of the characteristics of real-world brain networks.

The proposed novel generative model, NGPA, integrates the stochastic preferential attachment model with the exponential geometrical model, incorporating an intrinsic spatial scale denoted as $r_0$. By comparing the results of our model with real data, we determined that $r_0 \approx 3.5 mm$ within the structural human connectome. Our research further supports the significance of metric aspects in brain architecture. To some extent, the NGPA model enhances wiring efficiency and information processing efficiency simultaneously. The model reproduces several key network properties, although it does not properly capture the hyperbolic embedding properties or the $IPR_2$ peak at $\lambda = -1$ (which can be reproduced by increasing $r_o$ by a factor of 2.5 to 3, suggesting a typical scale for interacting TENS).

One possible explanation for the failure of the NGPA model to produce a good hyperbolic embedding of connectomes may be the existence of a "rich club" of nodes in the brain network \cite{richclub}. These "rich club nodes" are densely interconnected and play a crucial role in providing the shortest paths in the network, creating a heterogeneity among nodes based on their significance in information transfer. At the same time, the formation of a "rich club" is not accounted for in the NGPA model, which may lead to a more even distribution of node participation in signal transmission, despite the structural diversity present in the network. This immediately leads to an increase in Gromov hyperbolicity, since it is a measure of "democracy" in nodes' participation in optimal paths \cite{democracy}.

In this work, we explored the possibility of using preferential attachment to create hubs and construct the structural core of a network. While this was partially achieved, as shown in Fig. \ref{fig:rc-hyp}, this attempt was not fully successful. One reason for this may be due to the dissortative nature of the Barabási-Albert model with $\alpha>1$, which is inherent in the network \cite{nlBA}. To improve the NGPA model, a promising approach would be to replace preferential attachment with a more physically plausible principle for creating node heterogeneity. One such approach would be creating a hub by nonlinearly expanding the network \cite{Bauer2017}.

Our current investigation expands on the analysis of the localization of structural connectome modes initiated in \cite{moretti2013griffiths, pospelov2019spectral}. The spectrum includes soft cluster modes and a bulk continuum. States corresponding to the soft modes are localized within clusters. In agreement with the previous arguments in \cite{pospelov2019spectral}, we observed nontrivial correlations between clusters, which explain the semi-Poissonian level spacing behavior in the cluster bands found there.

A notable observation from \cite{pospelov2019spectral} is the mixed Wigner-Poisson behavior of $P(s)$ in the continuum, suggesting that the brain operates near a critical regime with a mobility edge that distinguishes  localized and delocalized modes within the bulk of the spectrum. We investigated the inverse participation ratio ($\text{IPR}_2$) of individual modes and qualitatively supported the former observation concerning the presence of both localized and delocalized modes in the bulk. However, some subtleties complicate the situation. Localized states form a tiny band around $\lambda=0$ scar-like states in the adjacency matrix spectrum originating from non-ideal non-interacting  TENs. Some of these states correspond to the trivial TENs corresponding to localization at the ends of the graph leaves, while others are clear-cut localized states at $\lambda=-1$, corresponding to interacting TENs.The boundary between these localized states and the rest of the delocalized modes in the continuum is not sharp, so there is no strict mobility edge in the bulk as anticipated in \cite{pospelov2019spectral}.

The analysis of $\text{IPR}_q$ reveals one more remarkable feature of the structural connectome. It turns out that there are multiple fractal states distributed throughout the spectrum. Furthermore, we have found signs of weak multifractality due to the quadratic dependence of $D_q(q)$. The power-law behavior of the return probability we found is another indication of multifractal states. With a lot of reservations we conjecture that the structural connectome operates in the extended multifractal phase. Certainly, the possibility that the structural connectome operates in the extended non-ergodic phase near criticality deserves further analysis. We can expect that fractal and scar-like structures play specific roles in the propagation of excitations within the brain, with scar-like modes residing predominantly in the bulk part of the Laplacian spectrum. Specifically, if the initial state significantly overlaps with a scar-like state, long-lived oscillations at the corresponding frequency may emerge. The fractality of the modes in the bulk is believed to slow down decay processes due to a power-law relation for the return probability, closely related to the decay rate.

Exploring the information theory aspects of our model, including various entropic measures, as discussed in \cite{bianconi2021information}, would be intriguing. For example, recently, the relation between fractal properties of the spectrum and the entanglement entropy of two subsystems has been explored in \cite{de2020multifractality}. It has been argued that under some additional assumptions the entanglement entropy saturates at $D>1/2$ generalizing the well-known Page saturation of the subsystem entropy. In our case we have $D>1/2$ indeed and natural separation of the system into two hemispheres. It would be interesting to investigate this in more detail.


Special attention should be paid to the account of edge directions. Despite the fact that there are many bidirectional connections in the brain that play an important functional role, such as providing stable zero-lag synchronization between cortical nodes \cite{Gollo2014}, for a complete understanding, it is also important to consider the direction of information transmission along axonal tracts. For example, it has been shown that in-weights gradients in the cortex determine traveling wave direction \cite{Koller2024}. However, the analysis of directed brain networks involves dealing with asymmetric adjacency matrices and complex spectra, which makes mathematical analysis difficult and limits the possibilities for theoretical understanding. We intend to delve into these issues in forthcoming publications.

\section{Data and Code Availability}

Analysis of connectomes and the construction of model networks were performed in Python. Some parts of this research were completed using the NetworkX library \cite{networkx}. The data used can be found at \href{https://braingraph.org}{braingraph.org database} \cite{PIT}. The code for the analysis and scripts producing the figures are available from the corresponding author, N.P., upon request.

\begin{acknowledgements}
We are grateful to I. Cheryomushkin for his collaboration at the early stages of the project, as well as to K. Anokhin and A. Brazhe for their useful discussions. The study was supported by the Non-Commercial Foundation "INTELLECT" for Support of Science and Education.
A.G. thanks IHES, where the part of the work has been done, for the hospitality and support.
\end{acknowledgements}

\bibliography{references.bib}

\newpage
\begin{appendix}

\section{Visualization of Laplacian eigenvectors}
\label{appendix:vis}

\begin{figure}[ht]
\centering
\includegraphics[width=14cm]{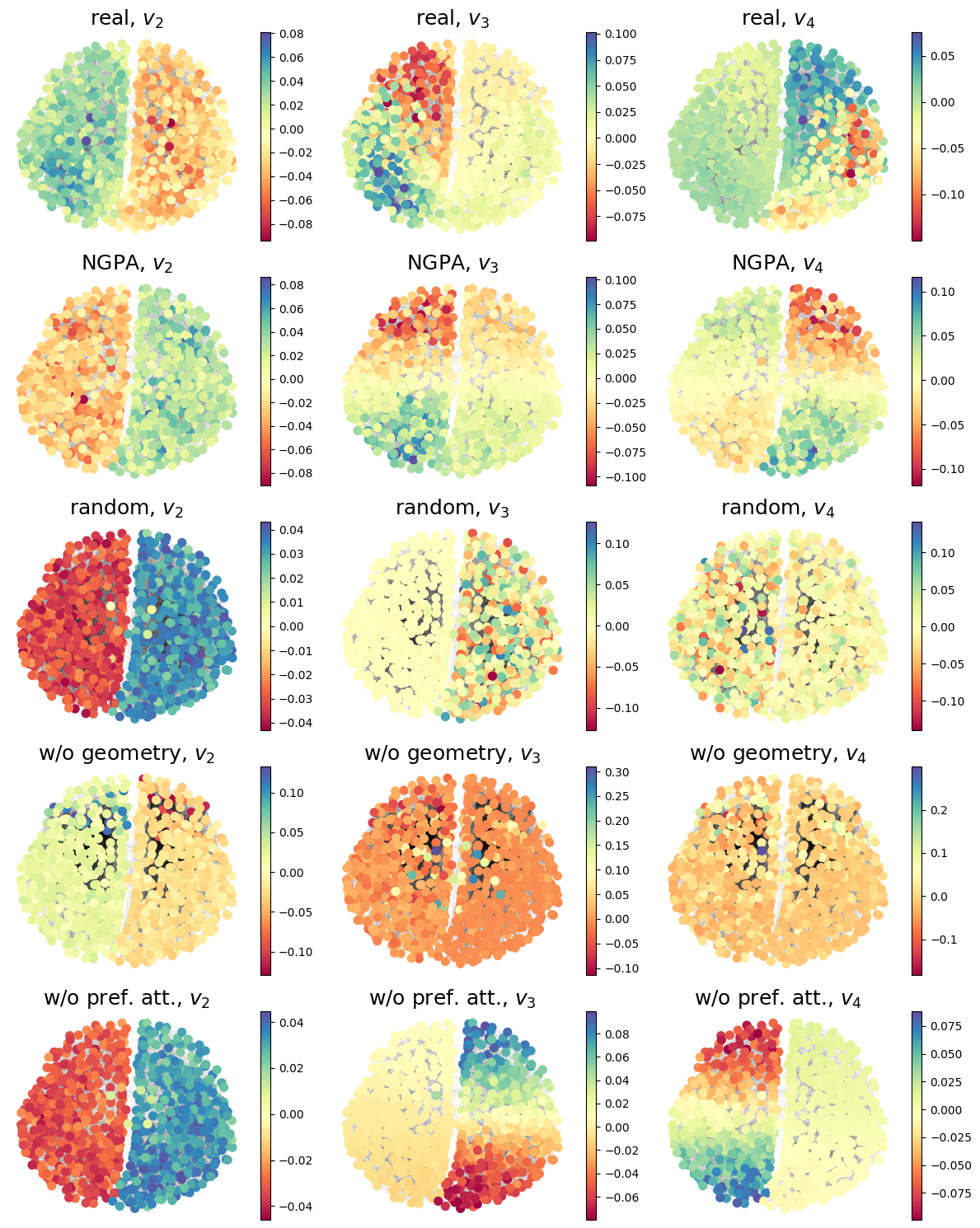}
\caption{Localization of $L^{\text{norm}}$ first eigenvectors of one real connectome and corresponding random models. Color encodes vector elements. Models shown: "random" = $M(0, 0)$, "w/o geometry" = $M(\alpha_{opt}, 0)$, "w/o pref.att." = $M(0, \beta_{opt}(0))$, "NGPA" = $M(\alpha_{opt}, \beta_{opt})$.}
\label{fig:brains}
\end{figure}

To better show the role of the first Laplacian eigenvectors in connectomes and artificial models, we visualized the first three of them ($v_2$, $v_3$ and $v_4$) in Fig. \ref{fig:brains}. The onstant eigenvector $v_1$ corresponding to $\lambda_1^\text{lap} = 0$ is omitted.

In all models except for $M(\alpha_{opt}, 0)$, the first non-trivial eigenvector $v_2$ divides the graph into two hemispheres, in full accordance with the idea of minimal cut in spectral clustering \cite{spclust}. The following eigenvectors $v_3$ and $v_4$ are localized in single hemispheres and divide them into approximately equal parts. For real connectomes and NGPA models, these parts are clearly localized, since due to the limited edge lengths, the optimal cut lies approximately at the middle of each hemisphere. This feature is especially pronounced in the model without preferential attachment, where the heterogeneity of node degrees does not violate the geometric separation of the hemispheres.

The effect of node heterogeneity is revealed in the behavior of eigenvectors in model $M(\alpha_{opt}, 0)$. The presence of hubs in the network distorts the geometric distribution of eigenvector elements, forcing them to concentrate on highly connected nodes. This breaks the usual symmetry between $v_3$ and $v_4$, and is the reason for the asymmetry in relative $\Delta\lambda_3$ and $\Delta\lambda_4$ for this model in Figure \ref{fig:lambdas}.

\section{Multifractality analysis}
\label{appendix:mfr}

For convenience, in this section we use the relative spectrum

\begin{equation}
\mu_i(\lambda_i) = \frac{i}{N} , i \in [1,N]
\label{eq:norm_lambda}
\end{equation}

When analyzing structural connectomes for the multifractality of eigenstates, we encountered several technical challenges. First, in contrast to artificial systems, working with brain networks means that we do not have the ability to generate new data quickly. Instead, we must work with a limited set of previously collected networks (\cite{PIT}). Second, even the highest-resolution networks are significantly smaller than the systems in which multifractal behavior is typically studied (see, for example, \cite{multifract}). Finally, the high level of individual variability in brain networks makes it difficult to apply simple heuristics for finding corresponding eigenmodes in systems of different sizes. For example, it is not possible to simply take an eigenvector that corresponds to the same normalized eigenvalue $\mu_i$.

To find the corresponding eigenstates in networks of different sizes, we used the following procedure:
\begin{itemize}
    \item Each eigenvector $v_j$ was projected onto an 87-dimensional "anatomical space" in which each anatomical area of the brain (region of interest, ROI) was represented by a separate dimension. The elements of this vector were calculated as $a_{i} = \sum_k(v_{j}^2)_k, k \in ROI_i$. In other words, $a_{i}$ contained aggregated ROI-related probabilities induced by the eigenvector $v_j$, summed over all anatomical subdivisions of this ROI on a given connectome resolution.

    \item The Jensen-Shannon divergence (JSD) was calculated between these "anatomical" vectors, as a measure of the information dissimilarity between two probability distributions over the ROIs. In the subsequent analysis, only groups of vectors with JSD values of $\leq 0.4$ were considered (since multifractal properties were analyzed using data at 4 scales, a total of $C_4^2=6$ JSD values were calculated).

    \item Additionally, we imposed a restriction on the relative positions of the eigenmodes in the spectrum: the relative eigenvalues at adjacent scales $S$ and $S+1$ should be close to each other: $|\mu_i^{S} - \mu_i^{S+1}| < 0.05$.

For the groups of eigenmodes selected in this way, the function $\text{IPR}_q(N)$ was calculated for different values of $q$. Only eigenmodes with a good linear fit to this function on a log-log plot ($R^2 > 0.95$ for every $q$) were considered for multifractality analysis.
    
\end{itemize}

\end{appendix}

\end{document}